\documentstyle[epsfig]{mn}
\begin{document}
\def\ltsima{$\; \buildrel < \over \sim \;$}
\def\simlt{\lower.5ex\hbox{\ltsima}}
\def\gtsima{$\; \buildrel > \over \sim \;$}
\def\simgt{\lower.5ex\hbox{\gtsima}}

\title[The properties of the absorbing and line emitting material in IGR J16318-4848 ]
{The properties of the absorbing and line emitting material in IGR J16318-4848}

\author[Giorgio Matt and Matteo Guainazzi]
{Giorgio Matt$^1$ and Matteo Guainazzi$^2$\\ ~ \\
$^1$Dipartimento di Fisica, Universit\'a degli Studi ``Roma 
Tre'', Via della Vasca Navale 84, I--00146 Roma, Italy \\
$^2$XMM-Newton Science Operation Center, RSSD of ESA, VILSPA, Apartado 50727, E-28080 
Madrid, Spain\\
}

\maketitle
\begin{abstract}
We have performed a detailed analysis of the XMM--Newton observation of IGR~J16318-4848, 
to study the properties of the matter responsible for the
obscuration and for the emission of Fe and Ni lines. 
Even if the line of sight material has a column density of about 2$\times10^{24}$ cm$^{-2}$,
from the Fe K$\alpha$ line EW and Compton Shoulder we argue that
the matter should have an average column density of a few $\times10^{23}$ cm$^{-2}$, along with
a covering factor of about 0.1--0.2. The 
iron K$\alpha$ line varies on time scales as short as 1000 s, implying
a size of the emitting region smaller than about 3$\times10^{13}$ cm. 
The flux of the line roughly follows the variations of the continuum, but not exactly,
suggesting a variation of the geometrical properties of the emitting region on similar
time scales. 
\end{abstract}


\section{Introduction}

IGR~J16318-4818 was discovered by
the ISGRI detector of the IBIS instrument onboard the INTEGRAL satellite
(Corvoisier et al. 2003) on January 29, 2003, with a 15--40~keV flux
of 50--100~mCrab. It was initially
interpreted as a new "transient" X-ray source. However, a
reanalysis of archival ASCA data revealed the presence of a source whose position was
coincident with that of IGR~J16318-4818 (Murakami et al. 2003), and with a
2--10~keV observed flux of about $4 \times 10^{-11}$~erg~cm$^{-2}$~s$^{-1}$.

The INTEGRAL discovery prompted a Target of
Opportunity observation with XMM-Newton
(Jansen et al. 2001) on February 10, 2003. The EPIC spectra unveiled
a variable and strongly absorbed source, with evidence for strong emission
lines (Schartel et al. 2003). This is in agreement with the ASCA
source, in which Revnivtsev et al. (2003) suggested a column
density $> 4 \times 10^{23}$~cm$^{-2}$. Preliminary
spectral fits (de Plaa et al. 2003) indeed suggested that
during the XMM-Newton observation IGR~J16318-4818 was
obscured by a Compton-thick absorber [$N_H = (1.66 \pm 0.16) \times
10^{24}$~cm$^{-2}$; all errors quoted hereinafter
are at the 90\% confidence level for one interesting
parameter]. The emission complex could be
resolved in three lines, with centroid energy of
$6.410 \pm 0.003$~keV, $7.09 \pm 0.02$~keV and $7.47 \pm 0.02$~keV.

In this paper we exploit the unprecedented sensitivity of the
EPIC detectors above 5~keV (where fluorescent
transition of Fe and Ni are concentrated) to characterize the
physical properties of the matter responsible for the
obscuration of the X-ray source and the associated reprocessing
features. To achieve this goal, we compare the spectral fit 
results (presented and discussed in Sect.~3.1 of this Letter)
with Monte-Carlo simulations of the X-ray transmission by neutral
matter (Matt et al. 1999; Matt 2002; see Sect.~3.2), originally developed
in the context of obscured AGN. We demonstrate in this paper that a proper
modeling of the scattering effect is crucial to correctly
identify the properties of the absorber, as standard fits with
approximate model can grossly overestimate the actual absorbing
column density and covering factor of the obscuring matter.
Study of the spectral variability during the XMM-Newton
observation allowed us to set stringent constraints on
the geometry of the absorbing matter (see Sect.~3.3).
Our findings and conclusions are summarized in Sect.~4.

\section{Data reduction}

In this paper, we discuss the pn data
only (Str\"uder et al. 2001), as this instrument seems to be the best
calibrated among the EPIC cameras
(see, for instance, the discussion in Molendi et al. 2003), at least
in their highest energy band.

Data were retrieved from the XMM-Newton Target of Opportunity public WEB page
as Observation Data Files (ODF), and reprocessed with SAS v.5.4.1,
using a calibration index file corresponding to the most updated calibration
files available at February 21, 2003. Spectra were extracted from a
circular region around the source centroid
($\alpha_{2000} = 16^h31^m48^s.6$,
$\delta_{2000} = -48^o49 \arcmin 00 \arcsec$; Schartel et al. 2003)
with a 52$\arcsec$ radius. As the extraction region encompassed two
CCDs, background spectra were extracted from a nearby source-free region,
having the same area ratio between the two chips as the source extraction
region. We verified that the results are not substantially different,
if the source and background pn spectra
are extracted from a smaller region (11$\arcsec$
radius), fully comprised in the same CCD. The spectra were accumulated
using single- and double-events. Pile-up is negligible, as well
as the contribution of the background, lower than 0.1\% in the 5--15~keV
energy band.
Nevertheless, a short interval of
about 1.25~ks duration, where the background count rate
(calculated on the single-event light curve extracted above 10~keV)
exceeded 1~$s^{-1}$, was removed from the scientific product accumulation.
The background is otherwise quite stable.
The fits described in this Letter were performed with XSPEC v.11.1.

\section{Data analysis and results}

\subsection{Time integrated spectrum: summary of spectral results}

As already noted by de Plaa et al. (2003), the XMM-Newton
spectra are characterized by a heavily absorbed continuum and 3 emission lines.
The lines are most naturally interpreted as  the Fe K$\alpha$
and K$\beta$ and the Ni K$\alpha$.
We therefore fitted the spectrum (in the 5-13 keV
energy band, as below 5 keV a small excess emission is present, see below) with the simplest
possible model: an absorbed power law plus three (unabsorbed) narrow
({\it i.e.} intrinsic width, $\sigma$, fixed to 1 eV) Gaussian lines. 
The photoelectric absorption model {\sc varabs} has been used with element
abundances from Anders \& Grevesse (1989) and photoelectric cross sections from
Balucinska-Church \& McCammon (1992). We left the iron abundance free to vary independently
of the other elements. As the absorbing matter results to be 
Compton--thick (see below), we included also Thompson absorption (model {\sc cabs}).
It is worth noting that this model, ignoring the scattering of photons,  is, 
strictly speaking,  valid only for absorbing
matter along the line of sight with a negligible covering 
factor (see below for a discussion), 


The fit is reasonably good ($\chi^2$=99.1/64 d.o.f.), but residuals around the iron 
K$\alpha$ line are visible, most likely due to the Compton Shoulder (CS), 
as already observed in the reflection spectrum of the Circinus Galaxy (Bianchi et al. 2002;
Molendi et al. 2003), and expected on theoretical ground (see Matt 2002 and references
therein). Modeling for simplicity
the Compton Shoulder with a Gaussian
with centroid energy fixed to 6.3 keV, and
$\sigma$ fixed to 50 eV, a significant improvement
is found ($\chi^2$=80.9/63 d.o.f., corresponding to 99.96\% confidence level). In the
following we will refer to the complete model (absorbed power-law plus
4 Gaussian emission lines) as the {\sl baseline model}. The best
fit results are summarized in Table~\ref{bestfit}. The observed 2-10 keV flux is
6.7$\times$10$^{-12}$ erg cm$^{-1}$ s$^{-1}$. The flux corrected for absorption is
instead 1.1$\times$10$^{-9}$ erg cm$^{-1}$ s$^{-1}$, corresponding to a luminosity
of 1.3$\times10^{37} d^2_{10}$ erg s$^{-1}$, where $d_{10}$ is the distance to the source
in units of 10 kpc. 

Given the large column density of the line--of--sight absorber, if the covering factor of the 
absorbing matter is large, a significant contribution from photons scattered towards
the line of sight is expected, as discussed in Matt et al. (1999). 
We therefore fitted the spectrum with 
the Monte-Carlo model described in that paper. The fit is completely unacceptable.
This may be due to the fact that the model has been calculated adopting the
Morrison \& McCammon (1983) cross sections, which made use of the Anders \& Ebihara (1982)
element abundances. In this set the iron abundance is about 0.7 times that
given by Anders \& Grevesse (1983), while from Table~\ref{bestfit} it seems that the
actual iron abundance (mainly derived in the fit from the depth of the iron edge)
in the absorbing material is closer to the one given by the latter authors.
Moreover, the Matt et al. (1999)
model assumed spherical geometry and homogeneous matter, 
while the covering factor may be significantly smaller than one and the average
column density smaller than that on the line--of--sight (see below).

\begin{figure}
\epsfig{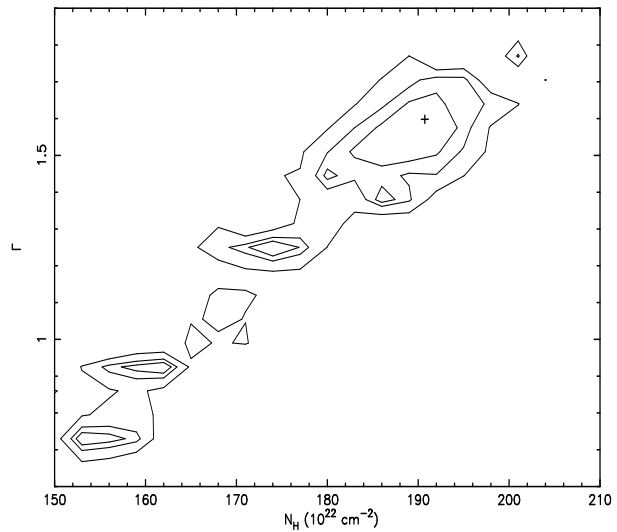}
\caption{The $\Gamma$-N$_{\rm H}$ contour plot. The contours correspond to 
$\Delta\chi^2$=2.3, 4.61 and 9.21.}
\label{gamma_nh}
\end{figure}

Coming back to the results summarized in Table~\ref{bestfit}, we must first of all remark
that the rather small errors on $\Gamma$ and N$_{\rm H}$ are somewhat misleading. In fact,
as it is clear from
the contour plot (Fig.~\ref{gamma_nh}), the $\chi^2$ distribution is
shallow and not very smooth. The calculated errors correspond to the $\Delta\chi^2$ in
the vicinity of the best fit, but other regions with $\Delta\chi^2$ less than 2.71 do exist.
We have verified that N$_{\rm H}$ and A$_{\rm Fe}$ are only slightly correlated each other.
To avoid problems in the estimate of the errors for the line parameters, 
we fixed $\Gamma$ and N$_{\rm H}$ to their best fit values. This does not affect very much
the errors on the Fe and Ni K$\alpha$ lines, but stabilizes the values for 
the Fe K$\beta$, which otherwise would be very difficult to find, due to its proximity to
the iron edge.  

\subsection{The physical properties of the absorber}

The energies of the Fe and Ni K$\alpha$ lines correspond to neutral or low ionized 
atoms (House 1969). On the contrary, the Fe K$\beta$ centroid energy is significantly
larger than expected. This cannot be due to high ionization, not only because it does not agree
with the  K$\alpha$ energy, but also because for significantly ionized matter the  K$\beta$
line becomes much fainter, to disappear completely for Fe {\sc xvii} or more, when no M
electrons remain. Instead the observed  K$\beta$/K$\alpha$ ratio
($ 0.20 \pm^{0.02}_{0.03}$) is slightly larger
than expected for neutral iron (see the
discussion in Molendi et al. 2003, where however the matter was seen in reflection,
not transmission; the arguments given there must therefore be taken with caution, until
proper calculations will be available).
It should however be noted that, given the proximity to the iron edge, the 
parameters of the K$\beta$ line are necessarily difficult to estimate, and may suffer from
a too simple modeling  of the edge (Palmeri et al. 2002). 

The Ni to Fe K$\alpha$ line ratio is about 6\%, suggesting a possible Ni overabundance
(see again the discussion in Molendi et al. 2003, with the same caution given above).
In Table~\ref{bestfit}, the EW of the lines with respect to the unabsorbed
continuum (to make easier the comparison with the expected value for the iron
K$\alpha$ presented in Matt 2002), are also given.
The value reported in Fig.~1 of Matt (2002) are
for Morrison \& McCammon (1983) abundances, and a power law index of 2. 
We therefore calculated the expected Fe K$\alpha$
properties using the code described in Matt (2002), adopting 
the Anders \& Grevesse abundances and
the best fit value for $\Gamma$, N$_{\rm H}$, and A$_{\rm Fe}$. 
The expected value of the ratio, $f$,  between the Compton Shoulder and the line core is 0.44,
to be compared with a measured value of
$0.12 \pm^{0.04}_{0.05}$.  $f$ does not depend much on the geometry, but rather
on the column density (Matt 2002). The observed value would correspond to a column density
of a few $\times10^{23}$ cm$^{-2}$, for which values around 100 eV of the EW 
are expected (a value of about 20 eV is instead expected for 1.9$\times10^{24}$ cm$^{-2}$).
The observed EW is instead about 13 eV. It is then possible that
the matter is very inhomogeneous, with a denser blob 
just on the line of sight (which is what the fit can measure)
but an average optical depth an order of magnitude less, and 
a covering factor, taking into account the uncertainties on the power law
index, of about 0.1--0.2. The lower (with respect to the line of sight) average
column density, along with the relative small covering factor, would explain
the failure of the Matt et al. (1999) model in fitting the data.

\begin{figure}
\epsfig{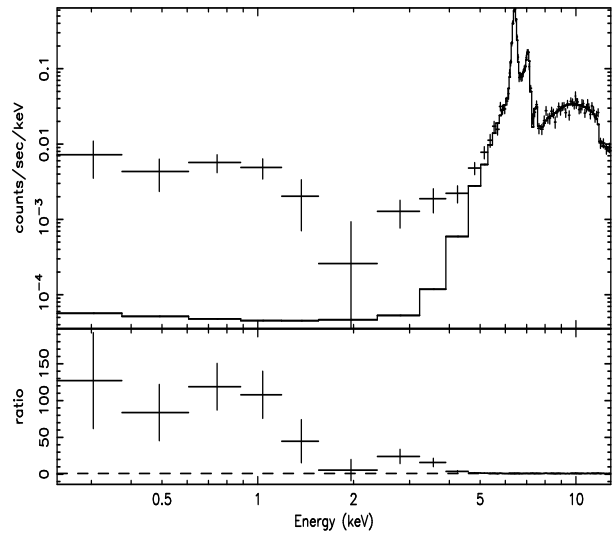}
\caption{The overall 0.3-13 keV time integrated spectrum fitted with the baseline model.}
\label{allsp}
\end{figure}

Because there is evidence that the absorbing material has a covering factor less than 1,
part of the X--ray illuminated surface should be directly visible, producing a  
Compton reflection component (George \& Fabian 1991; Matt, Perola \& Piro 1991) as
commonly observed in Compton--thick AGN (Matt et al. 2000 and references therein).
As discussed above, the average column density is possibly as low as a few 
$\times10^{23}$ cm$^{-2}$; however, below the iron line energy the reflection component
for this column density is very similar to that for Compton--thick matter
(Matt et al. 2003). This component could therefore account for
 the excess emission below 5 keV (see Fig.~\ref{allsp}, where
the whole 0.3-13 keV spectrum is shown, after being fitted with the baseline model),
and down to about 2 keV (the further excess at lower energies should have a different
origin). Fitting the 2-13 keV spectrum with the
baseline model gives $\chi^2$=104.9/68 d.o.f., and a very flat ($\Gamma$=0.6) power law. 
The addition of a pure Compton reflection component  (with the photon index linked to
that of the absorbed power law, and fixed to 1.6) improves the fit significantly, giving
$\chi^2$=79.9/68 d.o.f.. The value of $R$, 0.003, implies that the visible part of the
illuminated matter is very small ($R$ is equal to 1 for 2$\pi$ visible solid angle, i.e.
a covering factor of 0.5). 
As the source lies on the Galactic plane, absorption from interstellar matter is likely to
be significant. We then added an absorption component, but the fit does not
significantly improve 
($\chi^2$=77.9/68 d.o.f.). The best-fit face value
for the column density of this further absorber is about 5$\times10^{22}$
cm$^{-2}$, and $R$=0.0045. (If this is indeed the column density of the
interstellar absorption,  the emission below 2 keV is likely
due to another, nearby confusing source.)  The other parameters are similar to those listed
in Table~\ref{bestfit}.
The iron line EW with respect to the reflection component is very large, $\sim$28 keV, implying
that almost all of the line is related to the transmitted component.
Of course, part
of the excess emission may be due to photons scattered in transmission rather than
reflection, i.e. escaping from the far side (with respect to the X-ray source) of the
obscuring matter.

The small value of $R$ may, at the first glance, appears 
rather surprising, given the value of the covering
factor deduced from the iron line EW and the CS (about 0.1). 
It may be explained if, e.g., the absorber has a
flat configuration (similar to the `torus' envisaged in Unification Models for AGN),
seen at high inclination.

\begin{table}
\caption{Best fit results for the baseline model. 
Equivalent widths are calculated against the unabsorbed
continuum.} 
\begin{tabular}{||l|c|||}
\hline
\hline
& \cr
$\Gamma$ & 1.60$^{+0.07}_{-0.11}$ \cr
N$_{\rm H}$ (10$^{24}$ cm$^{-2}$) & 1.91$^{+0.03}_{-0.04}$ \cr
A$_{\rm Fe}^{a}$ & 0.89$^{+0.04}_{-0.03}$ \cr
E (Fe K$\alpha$) [keV] & 6.401$^{+0.001}_{-0.001}$ \cr
F (Fe K$\alpha$) [10$^{-5}$ ph cm$^{-2}$ s$^{-1}$] & 14.8$^{+0.4}_{-0.6}$ \cr
EW (Fe K$\alpha$) [eV] & 13 \cr
E (Fe K$\beta$) [keV] & 7.099$^{+0.001}_{-0.006}$ \cr
F (Fe K$\beta$) [10$^{-5}$ ph cm$^{-2}$ s$^{-1}$] & 3.05$^{+0.33}_{-0.38}$ \cr
EW (FeK $\beta$) [eV] & 3 \cr
E (Ni K$\alpha$) [keV] & 7.45$^{+0.05}_{-0.02}$ \cr
F (Ni K$\alpha$) [10$^{-5}$ ph cm$^{-2}$ s$^{-1}$] & 0.85$^{+0.19}_{-0.21}$ \cr
EW (Ni K$\alpha$) [eV] & 1 \cr
F (Fe K$\alpha$ CS) [10$^{-5}$ ph cm$^{-2}$ s$^{-1}$] & 1.88$^{+0.59}_{-0.68}$ \cr
& \cr
\hline
\hline
\end{tabular}
~\par
$^{a}$ relative iron abundance in solar units (Anders \& Grevesse 1989) by number.
\label{bestfit}
\end{table}

\subsection{Time resolved spectra}

\begin{figure}
\epsfig{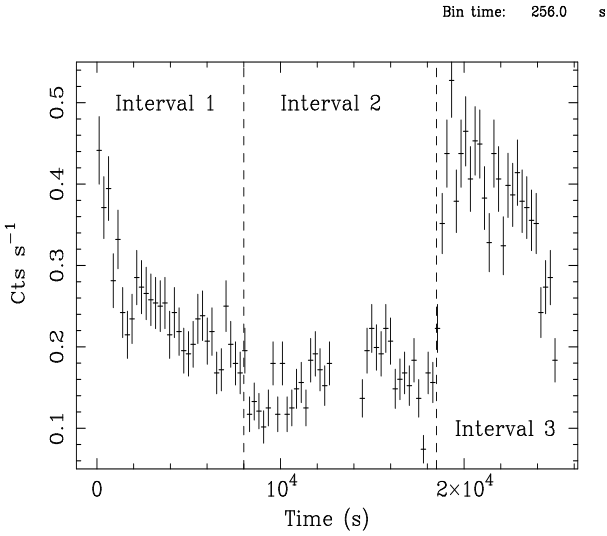}
\caption{pn light curve of the XMM-Newton observation of
IGRJ~16318-4848 (single- and double-events; 0.5--15~keV
energy band). The {\it vertical lines} indicate the time
intervals, where time-resolved spectra discussed in Sect.~3.3
were extracted}
\label{fig1_mg}
\end{figure}

\begin{figure}
\epsfig{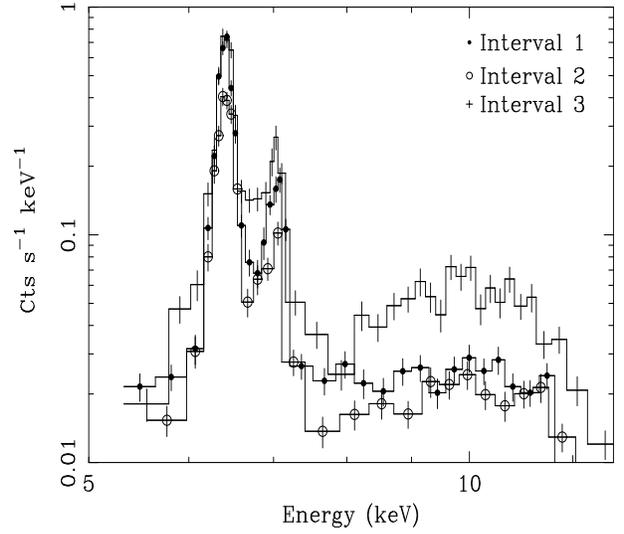}
\caption{pn count spectra for the time-resolved intervals
shown in Fig.~\ref{fig1_mg}}
\label{fig2_mg}
\end{figure}

IGR~J16318-4848 exhibits a complex variability pattern during the XMM-Newton
observation (see Fig.~\ref{fig1_mg}). The pn flux varied by a factor
of 2.5 during the first 8~ks, stabilized at a constant level of
$\simeq$0.15$s^{-1}$ (0.5-15~keV band) for the next 13~ks
and than underwent a sudden burst, brightening by a factor $\simeq$3
within $\simeq$500--1000~s. This behavior is associated with
spectral variability. In Fig.~\ref{fig2_mg} we show the count spectra
extracted in the 3 consecutive different time intervals highlighted
in Fig.~\ref{fig1_mg}: 0--8~ks, 8--18.5~ks, and 18.5--25~ks (elapsed
time) after the observation start. The exposure times corresponding to the
three spectra are 7.7, 7.3 and 5.1~ks, respectively. 

To test whether the flux variations are due to a change in the column density of the
absorber, or rather
in the properties of the primary emission, we performed three different
fits: with $\Gamma$ and the normalization of the power law kept fixed to the time
integrated best fit values (baseline model), 
and N$_{\rm H}$ free to vary (model 1); with only $\Gamma$ 
fixed and the normalization and N$_{\rm H}$ free (model 2); and with $\Gamma$ and
normalization free and  N$_{\rm H}$ fixed (model 3). For the sake of simplicity,
the fits have been performed in the 5-13 keV energy band. Hereinafter, all
fits have been performed with the
centroid energies of the emission lines
(which do not exhibit significant evidence for variations
among the time-resolved spectra) and the relative iron abundance 
kept fixed to the best fit values for the baseline model, as reported
in Table~\ref{bestfit}. The resulting $\chi^2$ are 
given in Table~\ref{chi2}. While for the 2nd interval all models are acceptable, 
for the 1st and 3rd intervals model 1 is unacceptable, while model 2 and 3 are both
acceptable. For model 2, the best fit values of
N$_{\rm H}$ are (see also Table~\ref{time_res}): 
1.76$^{+0.11}_{-0.10}$, 1.93$^{+0.15}_{-0.12}$
and  2.06$^{+0.09}_{-0.09}$.
For model 3, the best fit values of $\Gamma$ are: 1.85$\pm^{0.20}_{0.10}$,
1.55$\pm$0.19 and 1.30$\pm$0.18.

\begin{figure}
\epsfig{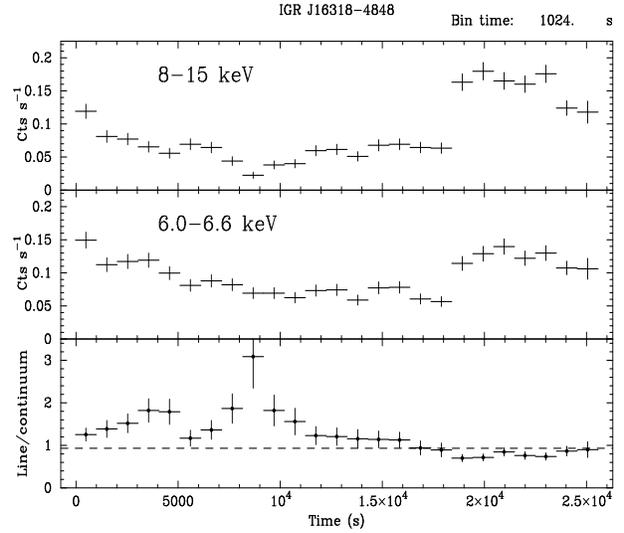}
\caption{The light curves in the 8-15 keV (upper panel) and 6.-6.6 keV (middle panel)
energy ranges.
In the latter band, the contribution from the iron K$\alpha$ line is dominant. In the lower
panel, the ratio of the two light curves is shown.} 
\label{fig3_mg}
\end{figure}

In Table~\ref{time_res} the best fit parameters for the absorbing column density and
the 8--15~keV and line fluxes 
(obtained with model 2; the results on the line fluxes are similar if model 3 is adopted)
are summarized. It is interesting to note that the fluxes of the lines are  correlated
with the flux of the continuum; the only exception
is the Compton shoulder, but the errors are pretty large.
This applies in particular for the spectra corresponding to Interval~2 (pre-burst)
and 3 (post-burst).
The ratios of the iron line fluxes between the post- and pre-burst phases
($1.8 \pm 0.2$ and $2.4 \pm^{0.8}_{1.3}$ for K$\alpha$ and
K$\beta$, respectively) are comparable with the average 8--15~keV
continuum flux ratio between the same intervals ($2.41 \pm 0.04$).
To further explore the relation between the emission line and the continuum fluxes,
in Fig.~\ref{fig3_mg} the light curves for the 8--15 keV (which is dominated by the 
continuum) and the 6--6.6 keV (dominated by the iron K$\alpha$ line) are shown. The 
line varies on time scales as short as 1000 s, implying that the size of the emitting
region cannot be larger than about 3$\times10^{13}$ cm, consistent
with a scenario in which the absorbing matter is e.g. due to the stream of material flowing
through the Lagrangian point in a Roche Lobe overflow, to eventually form an accretion disc.

The line flux follows the variations of the continuum, but not precisely, as it is clear
from the line--to--continuum flux ratio (lower panel of Fig.~\ref{fig3_mg}), suggesting that
on these time scales also the properties of the cold matter, i.e. the covering factor and/or
the average column density, change. 

\begin{table}
\caption{$\chi^2/d.o.f.$ for the three time intervals for different models.} 
\begin{tabular}{||l|c|c|c||}
\hline
\hline
& & & \cr
& 1st int. & 2nd int. & 3rd int. \cr
Count rate (5-13 keV) & 0.36 & 0.27 & 0.54 \cr
& & & \cr 
\hline
& & & \cr
$\Gamma$, Norm fixed & 56.2/35 & 24.4/21 & 88.8/34 \cr
$\Gamma$ fixed & 41.9/34 & 21.8/20 & 37.3/33 \cr
N$_{\rm H}$ fixed & 43.3/34 & 21.4/20 & 39.7/33 \cr
& & & \cr
\hline
\hline
\end{tabular}
\label{chi2}
\end{table}

\begin{table}
\caption{Best fit results for the three time intervals, using model 2 (see text for
details).} 
\begin{tabular}{||l|c|c|c||}
\hline
\hline
& & & \cr
& 1st int. & 2nd int. & 3rd int. \cr
& & & \cr
\hline
& & & \cr
N$_{\rm H}^a$  & 1.76$^{+0.11}_{-0.10}$ & 1.93$^{+0.15}_{-0.12}$
 & 2.06$^{+0.09}_{-0.09}$ \cr
F(8--15~keV)$^b$ & $2.38 \pm 0.04$ & $2.23 \pm 0.04$ & $5.38 \pm 0.03$ \cr
F (Fe K$\alpha$)$^c$  & 17.1$^{+1.2}_{-0.9}$ & 
10.1$^{+0.8}_{-1.0}$ &  18.4$^{+1.1}_{-1.7}$ \cr
F (Fe K$\beta$)$^c$  & 3.9$^{+0.6}_{-0.7}$ & 
 1.6$^{+0.7}_{-0.4}$ & 3.8$^{+0.8}_{-1.0}$ \cr
F (Ni K$\alpha$)$^c$ & 0.8$^{+0.5}_{-0.3}$  & 
0.3$^{+0.7}_{-0.3}$ & 1.5$^{+0.7}_{-0.6}$ \cr
F (Fe K$\alpha$ CS)$^c$ & 2.1$^{+0.8}_{-0.5}$ &
2.1$^{+0.6}_{-0.8}$ & 2.0$\pm$1.1 \cr
 & & & \cr
\hline
\hline
\end{tabular}
~\par
$^a$ in units of 10$^{24}$ cm$^{-2}$ \par
$^b$ in units of 10$^{-11}$~erg~cm$^{-2}$~s$^{-1}$ \par
$^c$ in units of 10$^{-5}$ ph cm$^{-2}$ s$^{-1}$
\label{time_res}
\end{table}

\section{Summary}

We have performed a detailed analysis of the XMM--Newton observation of IGR~J16318-4848, 
in order to characterize the properties of the matter responsible for the
obscuration along the line-of-sight and for the emission of Fe and Ni lines. 
Our results can be summarized as follows:

a) the line of sight material has a column density of about 2$\times10^{24}$ cm$^{-2}$;

b) from the value of the Fe K$\alpha$ line EW and Compton Shoulder, 
an average column density of a few $\times10^{23}$ cm$^{-2}$ (indicating
dishomogeneous or blobby material) and a 
covering factor of about 0.1--0.2 are estimated;

c) the small value of the Compton reflection component suggests a flat configuration of the
matter, with a large inclination angle;

d) the iron K$\alpha$ line varies on time scales as short as 1000 s, implying
a size of the emitting region not exceeding  $\sim3\times10^{13}$ cm. 
The flux of the line roughly follows the variations of the continuum, but not exactly,
suggesting a variation of the geometrical properties of the emitting region on similar
time scales. 

Finally, a few words on the putative ``transient" nature of this source.
A reanalysis of an archival
1994 ASCA observation (Murakami et al. 2003) points to a flux variation of a factor of just
a few in about 8.5~years. The comparison
between the flux measured by INTEGRAL in the 15--40~keV and by XMM-Newton in the 5--15~keV
is not conclusive, given the large uncertainties associated with the determination
of the intrinsic flux in such an absorbed source. Therefore, it cannot be ruled
out that the XMM-Newton (and ASCA) observations represent already its normal 
flux level. 

\section*{Acknowledgments}

This work is based on observations obtained with XMM-Newton,
an ESA science mission with instruments and contributions
directly funded by ESA Member States and the USA (NASA).
The XMM-Newton Science Operation Center is gratefully acknowledged
for having produced with unprecedented speed, and
made publicly available ODF for the observation described in
this paper. 

{}

\end{document}